\documentclass[3p,onecolumn]{elsarticle}

\usepackage{graphicx}
\usepackage{dcolumn}
\usepackage{pifont}
\usepackage{bm}
\usepackage{multirow}
\usepackage{amsmath}
\usepackage{hyperref}
\usepackage{booktabs}
\usepackage{float}
\usepackage{txfonts}
\usepackage[version=3]{mhchem}

\DeclareGraphicsExtensions{.pdf,.png,.jpg}

\usepackage[dvips]{color}
\definecolor{darkgreen}{rgb}{0,.6,0}

\begin{document}
\title{Propagation of Acoustical Wave in Finite Cylindrical Solid Bar surrounded by Semi-Infinite Porous Media saturated with Fluid}

\author[kimuniv-a]{Un-Son Ri}
\author[kimuniv-a]{Un-Gyong An}
\author[kimuniv-o]{Song-Jin Im}
\ead{}

\address[kimuniv-a]{Department of Acoustics, and}
\address[kimuniv-o]{Department of Optics, Faculty of Physics, Kim Il Sung University, Ryongnam-Dong, Taesong-District, Pyongyang, DPR Korea}

\begin{abstract}
We established the propagation equation of acoustical wave in media with the solid/porous media cylindrical boundary and obtained the analytic solution. We suggested the boundary condition on solid-porous media cylindrical boundary. Based on that, we introduced the dispersion equation, and constructed the algorithm to perform numerical calculation and analysis of dispersion equation.
\end{abstract}

\begin{keyword}
acoustic wave \sep porous media \sep fluid
\end{keyword}

\maketitle

\section{\label{sec:sec1}Introduction}
In the past, many research works have been performed in a fluid well surrounded by porous media. It is important for the understanding and quantitative interpretation of acoustic and seismic measurements in hydrocarbon wells, pipelines, as well as laboratory measurements to model wave propagation modes in cylindrical structures. For porous materials these waves are affected by material permeability. These effects can be analyzed using Biot’s equations of poroelasticity\cite{ref5}. For a porous cylinder with open-pore boundary conditions on its surface this was first done by Gardner\cite{ref6}, who derived the dispersion equation for extensional waves at low frequencies. For the full frequency range and closed boundary conditions the dispersion in a fluid-saturated cylinder was studied by Berryman\cite{ref7}.
This was done for the first few modes because conventional root finding becomes challenging for poroelastic media. An alternative approach to modeling wave propagation in circular structures was recently introduced by Adamou and Craster\cite{ref8} based on the spectral method. Karpfingeret al. \cite{ref9} extended the spectral method to axisymmetric waves for arbitrary fluid and solid layers. In this paper we considered the solution of dispersion equation of modeling wave in finite cylindrical solid bar surrounded by porous media saturated with fluid. Few literatures on sound propagation in solid bar surrounded by porous media is available\cite{ref1,ref2}. Today, many applications such as logging by solid bar and ultrasonic boring require theoretical studies on sound propagation in solid bar.

\section{\label{sec:sec2}Wave equation and its solution in solid bar and porous media}
In this paper, we assume that solid bar is homogenous axisymmetric one, porous media is the isotropic one whose porosity is $\phi$, and that linear dimension of porosity is much smaller than the wavelength.
Displacement of wave is given by \cite{ref1,ref2}
\begin{gather*}
u=\nabla\varphi+\nabla\times\nabla\times\psi \mathbf{e}_z
\end{gather*}
When the displacement oscillates harmonically, wave equation in solid bar may be written as
\begin{equation}
\nabla^2\varphi+k_p^2\varphi=0
\label{eq:eq1}
\end{equation}
\begin{equation}
\nabla^2\psi+k_p^2\psi=0
\label{eq:eq2}
\end{equation}
where $k_p^2=\omega^2/c_p^2$, $c_p^2=(\lambda+2\mu)/\rho$, $k_s^2=\omega^2/c_s^2$, $c_s^2=2\mu/\rho$, and $\lambda$, $\mu$ are Lame's constants and $\rho$ is the density of bar, $\nabla^2$ is laplacian, $\omega$ is frequency of wave.
In homogeneous and axial symmetrical bar the solutions of wave equation are
\begin{equation}
\varphi=AI_0(\overline{k_r}r)e^{ik_0z}
\label{eq:eq3}
\end{equation}
\begin{equation}
\psi=BI_0(\overline{k_r'}r)e^{ik_0z}
\label{eq:eq4}
\end{equation}
where $\overline{k_r}^2=k_0^2-k_p^2$, $\overline{k_r'}^2=k_0^2-k_s^2$, $k_0^2=\omega^2/c^2$, $I_n$ is modified Bessel function of the first kind of nth order, and $c$ is velocity of acoustical wave in axial direction. Equation of motion in porous media saturated as fluid is as follows\cite{ref3,ref4}
\begin{equation}
-N[\nabla\times[\nabla\times\mathbf{u}]]+(A+2N)\nabla\cdot\mathbf{u}+\overline{Q}\nabla(\nabla\cdot\mathbf{U})=\rho_{11}\ddot{\mathbf{u}}+\rho_{12}\ddot{\mathbf{U}}+b(\dot{\mathbf{u}}-\dot{\mathbf{U}})
\label{eq:eq5}
\end{equation}
\begin{equation}
Q\nabla(\nabla\cdot\mathbf{u})+\overline{R}\nabla(\nabla\cdot\mathbf{U})=\rho_{12}\ddot{\mathbf{u}}+\rho_{22}\ddot{\mathbf{U}}-b(\dot{\mathbf{u}}-\dot{\mathbf{U}})
\label{eq:eq6}
\end{equation}
where $\mathbf{U}$ and $\mathbf{u}$ are each displacement vector in solid and fluid, $\rho_{ij}$ and b are mass combination coefficient  and viscous coefficient respectively, which are given follows\cite{ref10}.
\begin{flalign*}
b(\omega)&=\beta^2H_1(\omega) \\
\rho_{22}(\omega)&=\frac{\beta^2H_2(\omega)}{\omega} \\
\rho_{12}(\omega)&=\beta\rho_f-\rho_{22}(\omega) \\
\rho_{11}(\omega)&=(1-\beta)\rho_s-\rho_{12}(\omega)
\end{flalign*}
Here
\begin{gather*}
H_1(\omega)+iH_2(\omega)=1/K(\omega)
\end{gather*}
$K(\omega)$ is Darcy coefficient, which is expressed as following\cite{ref11};
\begin{gather*}
K(\omega)=\frac{i\beta}{\omega\rho_f}\frac{J_2\left[ia\left(\frac{\displaystyle i\omega\rho_f}{\displaystyle \eta}\right)^{1/2}\right]}{J_0\left[ia\left(\frac{\displaystyle i\omega\rho_f}{\displaystyle \eta}\right)^{1/2}\right]}
\end{gather*}
$\rho_s$ and $\rho_f$ are densities of solid and fluid, $a$ is characteristic radius, $\eta$ is dynamic viscous coefficient of saturated fluid, $J_n$ is Bessel function of the first of order $n$, $\beta$ is porosity. $A$, $N$, $\overline{Q}$ and $\overline{R}$ are elastic constants related with Lame's constants, which are given as;
\begin{flalign*}
A&=\frac{(1-\beta)\left(1-\beta-\frac{\displaystyle K_m}{\displaystyle K_s}\right)K_s+\frac{\displaystyle K_s}{\displaystyle K_f}K_m}{1-\beta-\frac{\displaystyle K_m}{\displaystyle K_s}+\beta\frac{\displaystyle K_s}{\displaystyle K_f}}-\frac{2}{3}N \\
\overline{Q}&=\frac{1-\beta-\frac{\displaystyle K_m}{\displaystyle K_s}}{1-\beta-\frac{\displaystyle K_m}{\displaystyle K_s}+\beta\frac{\displaystyle K_s}{\displaystyle K_f}} \\
\overline{R}&=\frac{\beta^2-K_s}{1-\beta-\frac{\displaystyle K_m}{\displaystyle K_s}+\beta\frac{\displaystyle K_s}{\displaystyle K_f}} \\
N&=\mu_m
\end{flalign*}
where $K_s$, $K_m$, $K_f$ are bulk moduli of solid, skeleton of porous medium and fluid respectively, $\mu_m$ is shear modulus of skeleton of porous solid, which are given as follows
\begin{flalign*}
K_m&=(1-\beta)\rho_s\left(V_{l1}^2-4V_{s1}^2/3\right) \\
\mu_m&=(1-\beta)\rho_sv_{s1}^2 \\
K_f&=\rho_fV_{l2}^2
\end{flalign*}
$V_{l1}$, $V_{s1}$, $V_{l2}$ are velocities of longitudinal wave, shear wave in solid longitudinal wave in fluid of porous media respectively.
Let teh displacement vectors in solid, fluid within porous media be expressed as;
\begin{equation}
\mathbf{u}=\sum_{j=1}^2\nabla\varphi_j+\nabla\times\psi_1\mathbf{e}_\varphi
\label{eq:eq7}
\end{equation}
\begin{equation}
\mathbf{U}=\sum_{j=1}^2\xi_j(\omega)\nabla\varphi_j+\chi(\omega)\nabla\times\psi_1\mathbf{e}_\varphi
\label{eq:eq8}
\end{equation}
where
\begin{flalign*}
\xi_j(\omega)&=-\frac{1}{V_{lj}^2(\omega)}\frac{P\overline{R}-\overline{Q}^2}{\overline{Q}\gamma_{22}(\omega)-\overline{R}\gamma_{12}(\omega)}+\frac{\overline{R}\gamma_{11}-\overline{Q}\gamma_{12}(\omega)}{\overline{Q}\gamma_{22}-\overline{R}\gamma_{12}(\omega)} \\
\chi(\omega)&=-\gamma_{12}(\omega)/\gamma_{22}(\omega) \\
\gamma_{ii}(\omega)&=\rho_{ii}(\omega)-ib(\omega)/\omega \\
\gamma_{12}(\omega)&=\rho_{12}+ib(\omega)/\omega \\
p&=A+2N
\end{flalign*}
When acoustical field oscillates harmonically, we can introduce three Helmholtz's equation as follows.
\begin{equation}
\nabla^2\varphi_1+k_{p1}^2\varphi_1=0
\label{eq:eq9}
\end{equation}
\begin{equation}
\nabla^2\varphi_2+k_{p2}^2\varphi_2=0
\label{eq:eq10}
\end{equation}
\begin{equation}
\nabla^2(\psi_1\mathbf{e}_\varphi)+k_{s1}^2(\psi_1\mathbf{e}_\varphi)=0
\label{eq:eq11}
\end{equation}
where $k_{p1}^2=\omega^2/V_{l1}^2$, $K_{p2}^2=\omega^2/V_{s1}^2$, $V_{l1}$, $V_{l2}$ are velocities, $V_{s1}$ is velocity of rotational wave. Taking into account the boundary condition, solutions of above equations are
\begin{gather*}
\varphi_1=A_1K_0(K_{r1}R)e^{ik_0z} \\
\varphi_2=A_2K_0(K_{r2}R)e^{ik_0z} \\
\psi_1=B_1K_1(K_r'R)e^{ik_0z}
\end{gather*}
where $k_{r1}^2=k_0^2-k_{p1}^2$, $k_{r2}^2=k_0^2-k_{p2}^2$, $k_r'^2=k_0^2-k_{s1}^2$.

\section{\label{sec:sec3}solid-porous media cylindrical boundary condition}
Cylindrical boundary condition of solid-porous media and dispersion equation are
\begin{equation}
\begin{split}
(u_r)^I|_{r=a}&=(u_r)^{II} |_{r=a} \\
(u_z)^I|_{r=a}&=(u_r)^{II} |_{r=a} \\
(\sigma_{rr})^I|_{r=a}&=(\sigma_{rr})^{II} |_{r=a}+\sigma |_{r=a} \\
(\sigma_{rz})^I|_{r=a}&=(\sigma_{rz})^{II} |_{r=a} \\
(U_r)^{II}|_{r=a}&=(u_r) |_{r=a}
\label{eq:eq12}
\end{split}
\end{equation}
where superscripts I, II stand for solid and porous media. It is $S=-\beta P$\cite{ref3}, $\beta$ is porosity, $p$ is pressure in porous fluid, $u_r$, $u_z$ are displacements of solid in $r$ and $z$ directions, $\sigma_{rr}$, $\sigma_{rz}$ are components of stress, $U_r$ is displacement in $r$-direction in porous fluid.

\section{\label{sec:sec4}Dispersion equation}
We calculated the components of displacement and stress from solutions of equations in bar and porous media. The results are as follows.
\begin{flalign*}
(u_r)^I=&A\overline{k_r}I_1(k_rR)-Bik_0\overline{k_r'}I_1(\overline{k_r'}), \\
(u_z)^I=&-Aik_0I_0(\overline{k_r}R)-B\overline{k_r'}^2I_0(\overline{k_r'}R), \\
(\sigma_{rr})^I=&A\left\{\left[(\lambda+2\mu)\overline{k_r}^2-\lambda k_0^2\right]I_0(\overline{k_r}R)-2\mu\frac{\overline{k_r}}{R}I_1(\overline{k_r}R)\right\}+ \\
 &+B\left[-2\mu ik_0\overline{k_r'}^2 I_0(\overline{k_r'}R)+2\mu\frac{ik_0\overline{k_r'}}{R}I_1(\overline{k_r'}R)\right]\mathbf{e} \\
(u_r)^{II}=&-A_1k_{r1}K_1(k_{r1}v)-A_2k_{r2}K_1(k_{r2}R)-B_1ik_0K_1(k_r'R) \\
(U_r)^{II}=&-A_1\xi k_{r1}K_1(k_{r1}R)-A_2\xi_2k_{r2}K_1(k_{r2}R)-B_1\chi ik_0K_1(k_r'R) \\
(u_z)^{II}=&A_1ik_0K_0(k_{r1}R)+A_2ik_0K_0(k_{r2}R)-B_1k_r'K_0(k_r'R) \\
(\sigma_{rr})^{II}=&A_1\left[2Nk_{r1}^2K_0(k_{r1}R)+\frac{2Nk_{r1}}{R}K_1(k_{r1}R)+(A+Q\xi_1)(k_{r1}^2-k_0^2)K_0(k_{r1}R)\right]+ \\
&+A_2\left[2Nk_{r2}^2K_0(k_{r2}R)+\frac{2Nk_{r2}}{R}K_1(k_{r2}R)+(A+Q\xi_2)(k_{r2}^2-k_0^2)K_0(k_{r2}R)\right]+ \\
&+B_12N\left[ik_0k_r'K_0(k_r'R)+\frac{ik_0}{R}K_1(k_r'R)\right] \\
(\sigma)^{II}=&A_1(Q+R\xi_1)(k_{r1}^2-k_0^2)K_0(k_{r1}R)+A_2(Q+R\xi_2)(k_{r2}^2-k_0^2)K_0(k_{r2}R) \\
(\sigma_{rr})^{II}+\sigma=&A_1\left[2Nk_{r1}^2K_0(k_{r1}R)+\frac{2Nk_{r1}}{R}K_1(k_{r1}R)+(A+Q+Q\xi_1+R)(k_{r1}^2-k_0^2)K_0(k_{r1}R)\right]+ \\
&+A_2\left[2Nk_{r2}^2K_0(k_{r2}R)+\frac{2Nk_{r2}}{R}K_1(k_{r2}R)+(A+Q+Q\xi_2+R)(k_{r2}^2-k_0^2)K_0(k_{r2}R)\right]+ \\
&+B_12N\left[ik_0k_r'K_0(k_r'R)+\frac{ik_0}{R}K_1(k_r'R)\right] \\
(\sigma_{rz})^{II}=&-A_12Nik_0k_{r1}K_1(k_{r1}R)+A_22Nik_0(-k_{r2})K_1(k_{r2}R)+B_1N(k_r'^2+k_0^2)K_1(k_r'R) \\
(P)^{II}=&-\frac{\sigma}{\beta}=-A_1\frac{Q_1+R\xi_1}{\beta}(k_{r1}^2-k_0^2)K_0(k_{r1}R)-A_2\frac{Q+R\xi_2}{\beta}(k_{r2}^2-k_0^2)K_0(k_{r2}R)
\end{flalign*}
Substituting the displacements and stresses into boundary condition, we obtain following matrix equation,
\begin{equation}
\left(
\begin{array}{c c c c c}
	m_{11} & m_{12} & m_{13} & m_{14} & m_{15} \\
	m_{21} & m_{22} & m_{23} & m_{24} & m_{25} \\
	m_{31} & m_{32} & m_{33} & m_{34} & m_{35} \\
	m_{41} & m_{42} & m_{43} & m_{44} & m_{45} \\
	m_{51} & m_{52} & m_{53} & m_{54} & m_{55}
\end{array} \right)
\left(
\begin{array}{c}
A \\ B \\ A_1 \\ A_2 \\ B_1
\end{array} \right)=0,
\label{eq:eq13}
\end{equation}
where
\begin{flalign*}
m_{11}&=\overline{k_r}I_1(\overline{k_r}a) \\
m_{21}&=ik_0I_0(\overline{k_r}a) \\
m_{31}&=\left[(\lambda+2\mu)\overline{k_r}^2-\lambda k_0^2\right] I_0 (\overline{k_r}a)-2\mu\frac{\overline{k_r}}{a}I_1(\overline{k_r}a) \\
m_{41}&=2\mu i k_0 \overline{k_r}I_1(\overline{k_r}a) \\
m_{51}&=0 \\
m_{12}&=ik_0\overline{k_r'}I_1(\overline{k_r'}a) \\
m_{22}&=-\overline{k_r'}^2 I_0(\overline{k_r'}a) \\
m_{32}&=2\mu i k_0-\overline{k_r'}^2 I_0(\overline{k_r'}a)-2\mu\frac{ik_0-\overline{k_r'}}{a}I_1(\overline{k_r'}a) \\
m_{42}&=-\mu(k_0^2\overline{k_r'}+\overline{k_r'}^3)I_1(\overline{k_r'}a) \\
m_{52}&=0 \\
m_{13}&=k_{r1}K_1(k_{r1}a) \\
m_{23}&=-ik_0K_0(k_{r1}a) \\
m_{33}&=-2Nk_r^2K_0(k_{r1}a)-\frac{2Nk_{r1}}{a}K_1(k_{r1}a)-(A+Q+Q\xi_1+R\xi_1)(k_{r1}^2-k_0^2)k_0(k_{r1}a) \\
m_{43}&=2Nik_0k_{r1}K_1(k_{r1}a) \\
m_{53}&=(1-\xi_1)k_{r1}K_1(k_{r1}a) \\
m_{14}&=k_{r2}K_1(k_{r2}a) \\
m_{24}&=-ik_0K_0(k_{r2}a) \\
m_{34}&=-2Nk_{r2}^2K_0(k_{r2}a)-\frac{2Nk_{r2}}{a}K_1(k_{r2}a)-(A+Q+Q\xi_2+K\xi_2)(k_{r2}^2-k_0^2)K_0(k_{r2}a) \\
m_{44}&=2Nik_0k_{r2}K_1(k_{r2}a) \\
m_{54}&=(1-\xi_2)k_{r2}K_1(k_{r2}a) \\
m_{15}&=ik_0K_1(k_r'a) \\
m_{25}&=k_r'K_0(k_r'a) \\
m_{35}&=-2N\left(ik_0k_r'K_0(k_r'a)+\frac{ik_0}{a}K_1(k_r'a)\right) \\
m_{45}&=-N(k_r'^2+k_0^2)K_1(k_r'a) \\
m_{55}&=(1-\chi)ik_0K_1(k_r'a)
\end{flalign*}
In order that matrix equation has a nontrivial solutions, the determinant of the coefficient must be zero. Thus, the following dispersion equation must be satisfied;
\begin{gather*}
\left|
\begin{array}{c c c c c}
	m_{11} & m_{12} & m_{13} & m_{14} & m_{15} \\
	m_{21} & m_{22} & m_{23} & m_{24} & m_{25} \\
	m_{31} & m_{32} & m_{33} & m_{34} & m_{35} \\
	m_{41} & m_{42} & m_{43} & m_{44} & m_{45} \\
	m_{51} & m_{52} & m_{53} & m_{54} & m_{55}
\end{array} \right|=0
\end{gather*}

\section{\label{sec:sec5}Numerical calculation and interpretation of dispersion equation}
\begin{figure} [!ht]
	\begin{center}
		\includegraphics{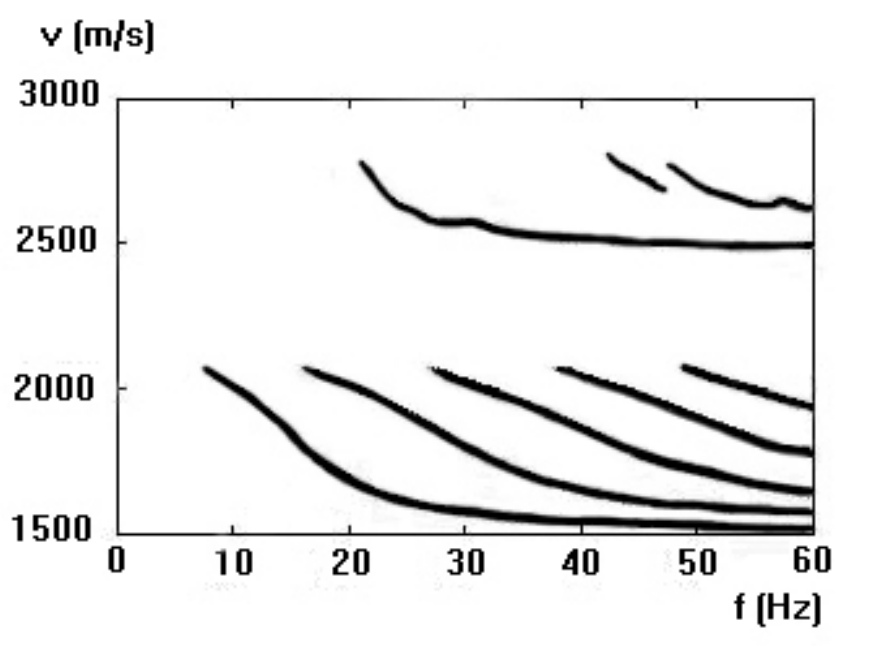}
		\caption{\label{fig:disp}Dispersion curve of velocity}
	\end{center}
\end{figure}

The dispersion equation is nonlinear equation of about $70^\mathsf{th}$, which makes the analytic solution impossible, so we must performed numerical evaluation. We used Newton method to do it. In case of the radius of bar $a=0.1m$, porosity $\phi=0.19$, viscous coefficient $\eta=0.01$, density of bar $\rho_1=1500kg/m^3$, velocity of longitudinal wave in bar $c_{p1}=2450m/s$, and velocity of tansversal wave in bar $c_{s1}=1500m/s$, density of solid, longitudinal and transpersal velocity in porous media $\rho_2=2650kg/m^3$, $c_{p2}=3670m/s$, $c_{s2}=2170m/s$ respectively, density of porous fluid $\rho_f=1000kg/m^3$, velocity in porous fluid $c_{p3}=1500m/s$, bulk modulus of porous solid $k_s=3.79\times10^{10}Pa$, the dispersion curve is shown in Figure~\ref{fig:disp}.
The dispersion curves have two groups. First group beginning from velocity of transversal wave in porous solid approach to velocity of transversal wave in bar, and second group from velocity of longitudinal wave in porous solid to velocity of longitudinal wave in bar. Imaginary parts of velocity of second group are very large, but in contrast, those one of second group are very small and ignored.
The first group of waves correspond to Transversal reference wave modes of solid bar and second group to longitudinal reference wave modes.
In case of $c_{s1}<c_{s2}$, transversal reference wave modes satisfy condition of total reflection.
But in contact of fluid of porous media with the bar, the condition of total reflection is not satisfied, therefore part of wave may be lost.
That's why imaginary part of wave number of undamped transversal reference wave appears in solid-to-solid boundary.
And the imaginary part of longitudinal reference wave of second group is large, so it can be shown that the condition of total reflection is not satisfied by existence of refraction transversal wave when longitudinal reference waves are reflected at boundary.
Therefore, longitudinal reference wave modes  become loss waves and transversal reference waves are fundamental part in a wave field propagating in bar and longitudinal reference waves  contribute little to complete wave field because of its damping property.

\section{\label{sec:con}Conculsion}
In this paper we established the equation of acoustical wave propagation in media with the solid-porous media cylindrical boundary and obtained its solution. We set solid-porous media cylindrical boundary condition to introduce the dispersion equation, and performed numerical evaluation of nonlinear equation of $70^\mathsf{th}$ degree to obtain phase velocity dispersion curves consisted of two groups, first of which corresponds to transversal reference wave mode, and second of which to longitudinal reference wave mode. The wave of transversal reference wave mode is undamped and that one of longitudinal reference wave mode is damped and ignored.

\bibliographystyle{plain}
\bibliography{Reference}

\end{document}